
\input phyzzx
\overfullrule=0pt
\def\n{\noindent}
\def\p{\partial}

\def\ee{\epsilon}
\def\ga{\alpha}
\def\gb{\beta }
\def\ap{\alpha '}

\def\sg{\sigma }

\def\IN{\relax{\rm I\kern-.18em N}}

\null
\rightline {UTTG-9-94}
\rightline {May 1994}

\title{ Asymptotic Level Density in Heterotic String Theory
and Rotating Black Holes
\foot{Work supported in part by NSF grants PHY89-17438,
PHY 9009850 and R.~A.~Welch Foundation.}}

\author{Jorge G. Russo \foot{
Address after September 1, 1994: Theory Division, CERN, CH-1211
Geneva 23, Switzerland} }
\address {Theory Group, Department of Physics, University of
Texas\break
Austin, TX 78712}
\author{Leonard Susskind}
\address {Department of Physics, Stanford University\break
Stanford, CA 94305-4060}

\abstract
We calculate the density of states with given mass and
spin in string theory and obtain asymptotic formulas.
We also compute the tree-level gyromagnetic couplings for
arbitrary physical states in the heterotic string theory.
These results are then applied to study whether fundamental strings can
consistently describe the microphysics of the
black hole horizon in the case of a general classical solution
characterized by mass, charge and angular momentum.

\bigskip
\endpage

\Ref\hagedorn {R. Hagedorn, Nuovo Cim. Suppl. 3 (1965) 147.}

\Ref\susskind {L. Susskind, {\it Some speculations about
black hole entropy in string theory}, Rutgers University preprint,
RU-93-44 (1993).}

\Ref\hoof{G. 't Hooft, Phys. Scripta T36 (1991) 247.}

\Ref\thoof{G. 't Hooft, Nucl. Phys. B256 (1985) 727.}

\Ref\uglum {L. Susskind and J. Uglum, {\it Black hole entropy in
canonical quantum gravity}, Stanford University preprint, SU-ITP-94-1
(1994).}



\Ref\carter{B. Carter, Phys. Rev. 174 (1968) 1559;
G.C. Debney, R.P. Kerr and A. Schild, J. Math. Phys. 10
(1969) 1842.}

\Ref\senbh {A. Sen, Phys. Rev. Lett. 69 (1992) 1006.}

\Ref\horowitz {J. Horne and G. Horowitz, Phys. Rev. D46 (1992) 1340.}

\Ref\ferrara {S. Ferrara, M. Porrati and V.L. Teledgi, Phys. Rev. D46,
(1992) 3529.}

\Ref\russo {J.G. Russo, {\it Thermal ensemble of string gas in a magnetic
field}, University of Texas preprint, UTTG-10-94 (1994).}

\Ref\sen {A. Sen, Phys. Rev. D32 (1985) 2102.}

\chapter {Introduction}

Massive excitations in the string spectrum constitute an
interesting and nevertheless scarcely explored sector in string theory.
Although the massless states suffice to describe all hitherto observed
physics, in processes involving planckian distances, such as e.g.
the big bang singularity or certain aspects of black hole physics,
the entire string spectrum is expected to play a role.
Highly-excited states are responsible of the ultraviolet finiteness of
string perturbation theory, and they lead to a number
of interesting effects. In particular, the level density
increases so rapidly with mass that the thermodynamical partition function
of a free string gas cannot be defined above a certain temperature
[\hagedorn ]. A significant part of the highly-excited states lies within
their Schwarzchild radii, and therefore they must be black holes.
States with sufficiently high-angular momentum have an average radius
larger than their Schwarzchild radii and thus they are not expected to be
black holes.

A statistical mechanical explanation
of the Bekenstein-Hawking entropy of black holes is unknown.
The black hole horizon behaves as a system
obeying the four thermodynamical laws, but their entropy was never explained
in terms of counting of quantum states. 't Hooft has stressed that
such an explanation may provide a clue on the fundamental nature of matter
[\hoof ].
Ordinary quantum field theory certainly does not give the correct
result for the entropy. It gives an ultraviolet divergent answer [\thoof ].
A recent proposal developed in ref. [\susskind ] and further studied in
ref. [\uglum ] entails a microscopic description of the
horizon in terms of configurations of fundamental strings in a
Rindler geometry.
This approach has provided an interesting explanation of the
Bekenstein-Hawking entropy, which is the essence of
the information paradoxes associated with Hawking radiation.

In this paper various properties of classical black holes will be compared
to a thermal average of fundamental strings in the heterotic string theory.
In sect 2 we will calculate the density level for
physical states in the string spectrum with a given angular momentum.
This will establish a precise
relation between the string energy, the angular momentum, and the ADM mass
of the configuration.

An intriguing property of rotating black holes in Einstein-Maxwell
theory is that they have gyromagnetic ratio equal to two, i.e.
the same as the tree-level $g$-factor of the electron  [\carter]. This
property holds true
for rotating black holes in the heterotic string theory [\senbh ],
but in other theories of gravity black holes may have $g\neq 2$, being,
in general, a function
of the black hole conserved quantum numbers (see, e.g., ref. [\horowitz ]).
We will derive the gyromagnetic couplings for
arbitrary physical states in the heterotic string theory
(in the case of the open string theory, this calculation
was done in ref. [\ferrara ]). Finally we will study the correspondence
with rotating black hole in heterotic string theory.

Related results, but in a different direction, are investigated
in ref. [\russo ].

\chapter {Density level of strings with given angular momentum}

In this section we will calculate an asymptotic formula for
the number of states  with a given mass and angular momentum.
We will consider the case of bosonic open strings, and then we
shall generalize our results to other cases.
To this purpose we modify the world-sheet Hamiltonian by adding
a term containing the angular momentum in the $z$ direction with a
Lagrange multiplier, i.e.
$$
H=\sum_{n=1}^\infty \sum_{i=1}^{D-2}
\ga_{-n}^i\ga_{n}^i+\lambda J\ ,\ \ \
\eqn\hamm
$$
where
$$
J=-i\sum_{n=1}^\infty {1\over n}\big( \ga_{-n}^1\ga_{n}^2
-\ga_{-n}^2\ga_{n}^1\big)\ .
$$
 The Hamiltonian can be diagonalized by writing
$$
\ga_n^1=\sqrt{n/2} \big(a_n+b_n\big)\ ,\ \
\ga_n^2=-i\sqrt{n/2} \big(a_n-b_n\big)\ ,\ \
$$
$$
\ga_{-n}^1=\sqrt{n/2} \big(a_n^{\dag} +b_n^{\dag} \big)\ ,\ \
\ga_{-n}^2=i\sqrt{n/2} \big(a_n^{\dag}-b_n^{\dag} \big)\ ,\ \
n=1,...\infty\ .
\eqn\oscil
$$
One has $[a_n,a_m^{\dag} ]=\delta_{nm}$ , $[b_n,b_m^{\dag} ]=\delta_{nm}$ .
The Hamiltonian takes the form
$$
H=\sum_{n=1}^\infty \bigg( \sum_{i=3}^{D-2}
\ga_{-n}^i\ga_{n}^i+ \big( n- \lambda \big) {a_n}^{\dag} a_n
+\big( n+ \lambda \big) {b_n}^{\dag} b_n \bigg)
\ .\ \ \
\eqn\hamdiag
$$
Let us now calculate the partition function,
$$
Z=\tr \big[ e^{-\beta H}\big] \ .
\eqn\parti
$$
We obtain
$$
Z=\prod _{n=1}^\infty \bigg[ \big( 1-w ^n\big) ^{-D+4}
\big( 1-cw^n\big) ^{-1}\big( 1-{w ^n\over c} \big)^{-1}\bigg ]\ ,
\eqn\partitio
$$
where $w \equiv e^{-\beta} $ and $c\equiv e^{\beta \lambda}$.

Note that $Z$ has poles at $\lambda=\pm 1, \pm 2,...$,
which can be traced back to the fact that $H$ will have negative
eigenvalues.

The partition function, eq. \partitio , can be expressed in terms of the
Jacobi $\theta $-function of the torus,
$$
\theta _1 \big( z|\tau )=2 f(q^2)q^{1/4} \sin (\pi z)
\prod_{n=1}^{\infty } \big( 1-2q^{2n} \cos (2\pi z)+ q^{4n}\big)\ ,
\eqn\jacobi
$$
where
$$
f(q^2)=\prod_{n=1}^\infty\big( 1-q^{2n}\big)=\bigg(
{1\over 2\pi q^{1/4}}{d\theta _1(z|\tau )\over dz}\bigg|_{z=0}
\bigg)^{1/3}=q^{-1/12}\eta (q^2 ) \
, \ \ \ \ q=e^{i\pi\tau }\ ,
$$
and $\eta (q^2)$ is the Dedekind eta function.
We find the following exact formula for the partition function
$$
Z(w,\lambda )=2{w^{1\over 8}\over f(w )^{D-5}}{\sin(\pi z )\over\
\theta _1(z|\tau )}\ ,\ \ \ z=-{i\gb \lambda\over 2\pi}\ ,\  \
\tau ={i\gb\over 2\pi}\ .
\eqn\exactparti
$$

In order to estimate the asymptotic density of states we use the modular
transformation property
$$
\theta_1(-{z\over\tau}|-{1\over \tau})=e^{i\pi\over 4}\sqrt{\tau }
\exp ({i\pi z^2\over \tau } ) \theta_1(z|\tau )\ .
\eqn\modtra
$$
Applying eq.\modtra\ to the partition function \exactparti\ we
obtain
$$
Z=\big(\beta/2\pi\big)^{D-4\over 2} w^{D-2\over 24}
e^{-{\gb\lambda ^2\over 2} } e^{a^2\over \gb }
{\sin(i\gb\lambda/2) f(e^{-{4\pi^2\over \gb}})^{4-D}\over
\sin(\pi\lambda ) g(\beta ,\lambda)}\ ,
\eqn\zasym
$$
where
$$
a\equiv \sqrt{D-2\over 6}\pi \ ,
$$
and
$$
g(\beta ,\lambda )\equiv \prod _{n=1}^\infty \big(
1-e^{i2\pi\lambda}e^{-{4\pi^2 \over \gb}n}\big)\big(
1-e^{-i2\pi\lambda}e^{-{4\pi^2 \over \gb}n}\big)\ .
$$
In the limit of high temperature $Z$ reduces to
$$
Z(\beta,\lambda )={\rm const.}\ \beta^{D-2\over 2} e^{a^2\over\beta
}{\lambda\over \sin (\pi\lambda )}\ .
\eqn\finale
$$

In order to extract the level density $d_{n,J}$ for states of
level $n$ and angular momentum $J$, it is convenient to expand $Z$ in
the following way
$$
Z(w,k)=\sum_{n,J} d_{n,J}w ^n e^{ik J}\ ,\ \ \ k =-i\beta\lambda \ .
\eqn\dnj
$$
Then $d_{n,J}$ can be projected out by Fourier integrating over $k$
and then integrating $w $ over a small circle around $w=0$ ,
$$
d_{n,J}={1\over 2\pi i} \oint {dw\over w^{n+1}} \int_{-\infty} ^\infty
{dk\over 2\pi} e^{-ik J}\ Z(w,k )\ .
\eqn\ddnj
$$
Interestingly, the integral over $k$ can be exactly carried out with
the result
$$
\int _{-\infty}^\infty dk e^{-i kJ }
{k \over \sinh (\pi k/\gb  ) }={\gb^2\over 2}
{1\over {\rm cosh}^2(\gb J/2)} \ .
\eqn\intek
$$
The integral over $w $ can be approximated by a saddle point evaluation.
Indeed, for large $n$ the integrand has a sharply defined saddle point at
$w$ given by the solution of the equation
$$
{a^2\over\gb^2}=n+1-J\ {\rm tanh}(J\gb /2)\ .
\eqn\saiddle
$$
If $J<<n$, that is, away from the Regge trajectories, the second term
on the right hand side can be ignored, and the solution is
$\gb \cong {a\over\sqrt{n+1}}$. If, on the other hand, $|J|=O(n)$, then
one can see from eq. \saiddle\ that
$|J|\gb >>1$ and therefore the stationary point is at
$$
\gb \cong {a\over\sqrt{n+1-|J|}}\ .
\eqn\sadbeta
$$
Since this solution also applies for $|J|<<n$ we can adopt it in the
general case. After performing the Gaussian integration,
what remains is
$$
d_{n,J}\cong {\rm const.}\ (n+1-|J|)^{-(D+3)/4} \exp
\big[{a(2(n+1)-|J|)\over \sqrt{n+1-|J|}}\big]
{1\over {\rm cosh}^2\big ({a J\over 2\sqrt{n+1-|J|} } \big) } \ .
\eqn\dos
$$
On the Regge trajectories $J=\pm n$ $d_{n,J}\to {\rm const.}$ which
can be normalized to 1.
Note that the number of states of level $n $ with zero angular
momentum $d_{n,0}$  differs from the total number of states of level $n$,
$d_n=\sum_J d_{n,J}$ only by a subleading factor of $1/\sqrt{n}$.
Integrating over $J$ we recover the familiar formula for $d_n$
(this can be done by integrating $J$ from $-n^{1-\ee}$ to $n^{1-\ee}$,
$\ee >0$, where $J$ can be neglected as compared to $n$).

The density of level as a function of the mass is ($J<<n$)
$$
\rho (m,J)={\rm const.} \ m^{-(D+1)/2} e^{m/T_H}
{1\over {\rm cosh}^2(a J/2\sqrt{\ap }m)} \ ,
\eqn\tres
$$
where $T_H$ is the Hagedorn temperature
$T_H=1/2a\sqrt{\ap}=1/4\pi \sqrt{\ap }$.

\chapter {Strings in a magnetic field and gyromagnetic coupling }

In this section we will obtain the tree-level
gyromagnetic coupling in three
different string theories, namely the bosonic
open string theory with U(1) Chan-Paton charges, a
bosonic string theory where U(1) charges arise from Kaluza-Klein
compactification, and the heterotic string theory.
\bigskip

\n {\bf A. Bosonic open string theory}

The world-sheet action in presence of an electromagnetic background
field is given by (to simplify the discussion we put a charge only at
one end of the string)
$$
S=-{1\over 4\pi \ap}\int d\tau d\sg
\p_\ga X^\mu\p^\ga X_\mu + \int d\tau d\sg q \delta(\sg )A_{\mu }(X)
\dot X^\mu \ .\ \
\eqn\actiopen
$$
In eq. \actiopen\ we have taken the sigma-model metric to be the Minkowski
metric. This is a valid string background (with vanishing beta-functions)
only up to terms of $O(F_{\mu\nu}^2)$, which can be
ignored for the purpose of deriving the gyromagnetic coupling which is
linear in $F_{\mu\nu}$.
In what follows all terms of $O(F_{\mu\nu}^2)$ will
be dropped.

The world-sheet Hamiltonian and the canonical momenta are given by
$$
H={1\over 4\pi \ap}\int_0^{\pi} d\sg \big[
(2\pi\ap)^2 \big( \Pi_\mu-q\delta(\sg )A_\mu(X)\big) ^2+{X'}^2\big]\ ,
\eqn\hami
$$
$$
\Pi_\mu={1\over 2\pi \ap}\p_\tau X_\mu + q \delta(\sg) A_\mu (X)\ .
\eqn\momopen
$$

Now let us consider the case in which only the component $F_{12}=-F_{21}$ is
different from zero, which for simplicity can be contemplated as
a constant magnetic field $F_{12}=$const.
The vector potential can be written as
$$
A_{\mu}=-{1\over 2}F_{\mu\nu}X^\nu\ .
\eqn\vepot
$$
In the case the magnetic field is not constant eq.\vepot\ provides the first
terms in an $\ap $ expansion. Successive terms are of higher order in
derivatives and will not affect the gyromagnetic coupling.

The boundary conditions at $\sg=0$ are
$$
X'_1(\tau,0)=2 \ap \pi q F_{12} \dot X^2(\tau,0)
\ ,\ \ \ X'_2(\tau,0)=2 \ap \pi q F_{21} \dot X^1(\tau,0)
\ ,\ \ \ \
$$
$$
X'_\mu (\tau,0)=0\ ,\ \ \ \mu\neq 1,2\ .
\eqn\boco
$$
In particular, we see that the magnetic field does not affect the boundary
conditions for the light-cone coordinates $X^\pm=(X^0\pm X^{D-1})/\sqrt{2}$,
so we can set, as usual, $X^+=x^+ +2\ap p^+\tau $.
By using eqs.\momopen\  and \hami \ we find for
the light-cone hamiltonian,
$$
2\ap p^+p^-={1\over 4\pi\ap }\int_0^{\pi} d\sg \big(
(2\pi\ap )^2\Pi_i^2+{X_i'}^2+(2\pi\ap )^2q\delta (\sg )F_{\mu\nu}
\Pi^\mu X^\nu \big)-1\ .
\eqn\lch
$$
Now it is easy to identify the gyromagnetic coupling
 for a physical state $|\Phi\rangle $
with mass $M$. We can measure the
shift in the energy due to the interaction between the spin and
the magnetic field.
We use $2p^-p^+=E^2 -p^{D-1}$ and expand the energy
in powers of $1/M$ (in the presence of a magnetic field we cannot choose
the rest frame because a combination of $p_1$ and $p_2$ does not commute
with the Hamiltonian. In any case, terms containing $\vec p$ do not provide
any contribution to the gyromagnetic coupling and are ignored). We obtain
$$
\langle \Phi| {\cal H}_{\rm mag} |\Phi \rangle
=\langle \Phi| -{q\pi \over 2 M}  F_{\mu\nu}
X^\mu (0,\tau  ) \Pi^\nu(0,\tau) |\Phi \rangle \ .
\eqn\hmag
$$
We can insert the
free expansion for the string coordinates since the $O(F_{\mu\nu})$
corrections will only give $O(F^2)$ contributions. We have
$$
X^\mu(0,\tau) =x^\mu + 2\ap p^\mu + i \sqrt{2\ap}\sum_{n\neq 0}{1\over
n}\ga _n^\mu
e^{-in\tau}\ ,\ \ \ \ \dot X^\mu(0,\tau )=2\ap p^\mu+
\sqrt{2\ap}\sum_{n\neq 0}\ga _n^\mu e^{-in\tau}\ .
\eqn\expan
$$
${\cal H}_{\rm mag} $ can be written as the sum of an orbital angular momentum
piece plus a purely spin part,
${\cal H}_{\rm mag}^o +{\cal H}_{\rm mag}^s $.
By inserting eqs. \expan\ in eq. \hmag\ and keeping the spin contribution
${\cal H}_{\rm mag}^s $ only we find
$$
\langle \Phi | {\cal H}_{\rm mag}^s |\Phi \rangle
=\langle \Phi|
-{q\over 2 M} S^{\mu\nu} F_{\mu\nu}|\Phi \rangle \ ,\ \ \
\eqn\unamas
$$
where
$$
S^{\mu\nu }=-i\sum _{n=1}^\infty {1\over n}\big(\ga_{-n}^\mu \ga_n^\nu
-\ga_{-n}^\nu \ga_n^\mu\big) .
$$
{}From the standard form of the magnetic coupling,
${\cal H}_{\rm mag} =- \vec \mu \cdot \vec  B $ we obtain that the magnetic
dipole moment is given by
$\vec \mu= {q\over  M} \vec J$, whereby we deduce that $g=2$
for all charged, spinning physical states in the bosonic open string theory.
This is in agreement with a previous calculation by Ferrara et al
[\ferrara ].
\bigskip

\n {\bf B. Bosonic string with U(1) Kaluza-Klein charges}
\bigskip

We begin by considering the sigma-model action for the bosonic
string in presence of a metric and antisymmetric field:
$$
S=-{1\over 4\pi \ap}\int d^2\sg \big[G_{AB}\eta^{\ga\gb }
\p_\ga X^A\p_\gb X^B + B_{AB}\ee^{\ga\gb} \p_\ga X^A\p_\gb X^B \big]\ ,
\eqn\action
$$
where $A,B=0,1,...,D-1$,
$\eta_{\ga\gb }$ has signature $(-,+)$, $\ee^{12}=-\ee^{21}=1$,
and $D=26$. Let us assume that
one dimension, $X^{D-1}\equiv \varphi $, is
compactified on $S^1$ and the metric and antisymmetric fields take the
following form:
$$
S=-{1\over 4\pi \ap}\int d^2\sg \big[\eta_{\mu\nu}
\p_\ga X^\mu\p^\ga X^\nu + \p_\ga \varphi \p^\ga \varphi +
2A_\mu^{\rm v} \p_\ga X^\mu\p^\ga\varphi+2A_\mu^{\rm a} \ee^{\ga\gb}
\p_\gb X^\mu\p_\ga \varphi \big] \ ,\
\eqn\kaction
$$
$$
\ \ \ \mu,\nu =0,1,...,D-2\ .
$$
Let us introduce $x^\pm=\tau \pm\sg $. The action \kaction\ becomes
$$
S={1\over\pi\ap}\int d^2\sg \big[ \p_+X^\mu\p_-X_\mu +\p_+\varphi \p_-\varphi
+(A^{\rm v}_\mu+A_\mu^{\rm a}) \p_-X^\mu\p_+\varphi
+(A^{\rm v}_\mu -A^{\rm a}_\mu )\p_+X^\mu\p_-\varphi \big]\ .\ \ \
\eqn\lefty
$$
In the particular case $A^{\rm a}_\mu=A^{\rm v}_\mu  $, the model will provide
a bosonic analog of heterotic string theory insofar as all the gauge quantum
numbers will originate from the left sector.

Let us derive the Hamiltonian corresponding to eq. \kaction . The canonical
momenta are given by
$$
\Pi_\mu={1\over 2\pi\ap }\big( \p_\tau X_\mu +A^{\rm v}_\mu\p_\tau\varphi
+A^{\rm a}_\mu\p_\sg\varphi\big)\ ,\ \
\eqn\mommu
$$
$$
\Pi_\varphi={1\over 2\pi\ap }\big( \p_\tau \varphi +A^{\rm v}_\mu\p_\tau X^\mu
-A^{\rm a}_\mu\p_\sg X^\mu \big)\ .\ \
\eqn\momenta
$$
The Hamiltonian is (as before we ignore $O(A^2)$ terms)
$$
H={1\over 4\pi\ap }\int_0^\pi d\sg \big[
(2\pi\ap )^2 \Pi^2_\mu+{X'}^2+(2\pi\ap )^2 \Pi_\varphi^2
+{\varphi '}^2-
$$
$$
2 A_\mu^{\rm v} \big( (2\pi\ap )^2 \Pi_\varphi\Pi^\mu-\varphi
'{X^\mu}'\big) - 2 A_\mu^{\rm a} (2\pi\ap )
\big( \varphi '\Pi^\mu-\Pi_\varphi {X^\mu}'\big) \ .
\eqn\hamilton
$$

Let us first discuss the particular case $A_\mu^{\rm a}=0$.
As in the previous subsection, let us consider the case in which
only the $F_{12}^{\rm v}=-F_{21}^{\rm v} $ is non-vanishing.
Then
the free-field equations of motion for the light-cone coordinates
$X^\pm $ are not affected and we can set $X^+=x^++2\ap p^+\tau $.

The term in the Hamiltonian containing the interaction
with the electromagnetic field is
$$
H_{\rm int}= {1\over 4\pi\ap } F_{\mu\nu}^{\rm v}\int_0^\pi d\sg
X^\nu \big( (2\pi\ap )^2 \Pi_\varphi\Pi^\mu-{X^\mu}'\varphi '\big)\ .
\eqn\nosecomo
$$

We note that there is an unusual term $\varphi 'F^{\rm v}_{\mu\nu} X^\nu
{X^\mu}' $
for states with non-vanishing winding number.

In order to identify the gyromagnetic coupling for a physical state
$|\Phi\rangle $
with mass $M$, again we use
$2p^-p^+=E^2 -p^{D-1}$ and expand the energy
in powers of $1/M$ . We obtain
$$
\langle \Phi| {\cal H}_{\rm mag} |\Phi \rangle
=\langle \Phi| {1\over 2\ap M} H_{\rm int}|\Phi \rangle \ .
\eqn\hmagk
$$
We can insert the free mode expansion for the string coordinates since
the $O(F^{\rm v}_{\mu\nu})$ corrections will only give $O(F^2)$ contributions.
Only the zero-mode part
of $\Pi_\varphi $ and $\varphi '$, representing charge and winding
number, respectively, will contribute to this linear
order in $F_{\mu\nu}^{\rm v}$ .
Thus we find
$$\eqalign {
\langle \Phi| {\cal H}_{\rm mag}^s |\Phi \rangle &=
-{Q\over 4 M}F^{\rm v}_{\mu\nu}S^{\mu\nu}
+{W\over 4 M}F^{\rm v}_{\mu\nu}(S^{\mu\nu}_L-S^{\mu\nu}_R)\cr
&\equiv -{Q\over 2 M}B_z \mu _z \cr }\ ,
\eqn\magnus
$$
where $Q$ and $W$ are respectively the U(1) charge and the winding number
given by
$$
Q={1\over 2\pi\ap } \int_0^\pi d\sg \dot \varphi \ \ ,\ \ \
W={1\over 2\pi\ap } \int_0^\pi d\sg \varphi '\ \ ,
\eqn\carga
$$
and $S^{\mu\nu }=S^{\mu\nu }_R+S^{\mu\nu }_L$,
$$
S^{\mu\nu }_R=-i\sum _{n=1}^\infty {1\over n}\big(\ga_{-n}^\mu \ga_n^\nu
-\ga_{-n}^\nu \ga_n^\mu\big)\ ,\ \
S^{\mu\nu }_L=-i\sum _{n=1}^\infty {1\over n}\big(\bar \ga_{-n}^\mu \bar
\ga_n^\nu-\bar \ga_{-n}^\nu \bar \ga_n^\mu\big) \ .
\eqn\espin
$$
{}From eq. \magnus \  one can see that the magnetic moment operator in general
does
not commute with $S^2$ and therefore they cannot be simultaneously
diagonalized. The gyromagnetic factor must be defined as the ratio
between expectation values. For states labelled by $S^2, S_z$ we have
$$
 \langle \Phi | \mu _z |\Phi \rangle =
g{Q\over 2M}  \langle \Phi| S_z |\Phi \rangle \ .
\eqn\gyrosin
$$
Thus in this theory physical states will have
$$
\langle g\rangle =1+{W\over Q} {\langle S_R-S_L\rangle \over
\langle S \rangle }\ .
\eqn\gyrokaluza
$$
It is interesting to note the physical state $\ga_{-1}^\mu |Q,W\rangle $
(and similarly for the state $\bar \ga_{-1}^\mu |Q,W\rangle $),
because it is subject to the Virasoro condition $N_R-N_L=QW=1$,
it will have $g=2$ only at the self-dual point where $Q=W$.

Let us now consider the case $A^{\rm a}_\mu=A^{\rm v}_\mu  $,
$A_\mu^L = {1\over 2}(A^{\rm v}_\mu+A^{\rm a}_\mu )\neq 0 $,
which, as mentioned above, is a
bosonic analog of the heterotic string theory.
The term in the light-cone Hamiltonian containing the interaction
with the electromagnetic field is
$$
H_{\rm int}= {1\over 4\pi\ap } F^L_{\mu\nu}\int_0^\pi d\sg
X^\nu \big( 2\pi\ap \Pi^\mu - {X^\mu}' \big)
\big(\varphi' + 2\pi\ap \Pi_\varphi \big)\ .
\eqn\hinter
$$
By following similar steps as before we find
$$
\langle \Phi| {\cal H}_{\rm mag}^s |\Phi \rangle =-
{Q_L\over  2 M}F_{\mu\nu}^L S_R^{\mu\nu}=-{Q_L\over M}B_z S_R^z\ ,
\eqn\maghet
$$
where
$$
Q_L={1\over 2\pi\ap } \int _0^\pi d\sg \big(\varphi'+ \dot \varphi \big)\ ,
\eqn\cargaleft
$$
and
$$
S_R^{\mu\nu}=-i\sum _{n=1}^\infty {1\over n}\big(
\ga_{-n}^\mu \ga_n^\nu
-\ga_{-n}^\nu \ga_n^\mu\big) \ .
\eqn\espinr
$$
{}From eq. \maghet\  we conclude that in this model
physical states have a magnetic dipole moment $\vec \mu = {Q_L\over M} \vec
S_R$.
The same result will appear in the case of the heterotic string theory
considered below.

\bigskip

\n {\bf C. Heterotic string theory}

The action of the heterotic string in presence of background
gauge fields is given by [\sen ]
$$
S=-{1\over 4\pi\ap }\int d^2\sg\big[ \p_\ga X^\mu\p^\ga X_\mu -
2i\psi_-^\mu\p_+\psi_{-\mu}-2i\lambda_+^r\p_-\lambda_+^r -
$$
$$
2 \lambda^r_+ (T^M)_{rs} \lambda^s _+ A^M_\mu \p_-X^\mu+
{i\over 2}\lambda^r_+ (T^M)_{rs} \lambda^s _+
F_{\mu\nu}^M\psi_-^\mu\psi_-^\nu \big]\ .
\eqn\acthet
$$
Again, $O(F^2)$ corrections to metric and other sigma-model backgrounds
will be ignored since they do not affect the gyromagnetic coupling.
The canonical momenta are given by
$$
2\pi\ap \Pi_\mu=\dot X_\mu+ \hat O^M A_\mu^M\ ,\ \ \
\hat O^M \equiv {1\over 2}\lambda^r_+ (T^M)_{rs} \lambda^s _+\ ,
$$
$$
2\pi\ap \Pi_\psi^\mu={i\over 2}\psi^\mu_-\ ,\ \ \
2\pi\ap \Pi_\lambda^r = {i\over 2} \lambda ^r_+\ .\ \ \
\eqn\canhet
$$
One finds for the Hamiltonian
$$
H={1\over 4\pi\ap}\int_0^\pi d\sg
\big[(2\pi\ap )^2 \Pi^2_\mu+{X'_\mu}^2-
i\psi^\mu_-\p_\sg\psi^\mu_- + i\lambda^r_+\p_\sg\lambda^r_+
$$
$$
-2 \hat O^M A^M_\mu \big( 2\pi\ap \Pi^\mu -{X^\mu}'\big)
+i \hat O^M F_{\mu\nu}^M\psi_-^\mu\psi_-^\nu\big]\ .
\eqn\hamhet
$$

Now we consider the particular background where only one of the
components, $F^1_{12}=-F^1_{21}$, is different from zero. The
light-cone gauge can then be fixed in the standard way by setting
$X^+=x^++2\ap p^+\tau $ and $\psi^+=0 $. Proceeding as in the previous
model, we obtain that the magnetic dipole moment for a physical state
$|\Phi\rangle $ with charge $q$ and mass $M$ is given by
$$
\mu ^z= {q\over M} S^z_R\ ,
\eqn\muhetero
$$
$$
q= \langle \Phi|{1\over 2\pi \ap}\int_0^\pi d\sg
{1\over 2}\lambda^r_+ (T^1)_{rs} \lambda^s _+|\Phi\rangle \ ,
\eqn\carica
$$
where $S^z_R$ is the right contribution to the angular momentum,
that is,
$$
S^z_R= - i\sum _{n=1}^\infty {1\over n}\big(
\ga_{-n}^1 \ga_n^2-\ga_{-n}^2\ga_n^1 \big)
-{i\over 2}[d_0^1,d_0^2] -
i \sum_{n=1}^\infty \big( d_{-n}^1
d_{n}^2 - d_{-n}^2 d_{n}^1\big) \ ,
\eqn\espinram
$$
in the Ramond sector, and the analog expression in the Neveu-Schwarz
sector. This is the result anticipated above.

Note that the massless
particles with gauge quantum numbers in the heterotic string theory
get all their
angular momentum from the right sector, that is $S_R^{\mu\nu}=S^{\mu\nu}$.
As a result, $\mu^z=q S^z/M$, and thus they have $g=2$ (a small mass $M$
is given by sitting infinitesimally away the self-dual point).

In the general case $g$ will be given by
$g=2\langle S_R^z\rangle/S^z$.
In the next section  we will be interested in a thermal ensemble of
heterotic string states with macroscopic masses ($M^2>>1/\ap $). In this
case we obtain the average gyromagnetic
factor of a thermal average of heterotic strings with a given mass,
charge and angular momentum.

\chapter {Correspondence with rotating black holes}

In ref. [\susskind ] (see also [\uglum ]) a statistical mechanical
explanation of the Bekenstein-Hawking entropy as counting of quantum states
was suggested. The proposal entails a description of the black hole
horizon in terms of strings moving in a Rindler geometry.
This study was carried out for the Schwarzchild black hole and a remarkable
correspondence was obtained. It is important to understand to what extent
a string description is feasible. So let us consider the most general
solution given by the rotating, charged black hole solution, that is,
the heterotic string analog of the Kerr-Newman black hole (see ref. [\senbh ]).
We shall compare our formulas for the level density
derived for a free string with the entropy of the rotating black hole
solution. It should be noted that the angular momentum is quantized and
it is also an adiabatic  invariant, and therefore it should not change if
the gravitational coupling were gradually increased in such a way the
initial free string  gravitational collapses.

In addition to the entropy, if  a charged and rotating black hole is
to be described in terms of a fundamental string, it should have the same
gyromagnetic
factor. The gyromagnetic factor of a black hole varies according to the theory
(see e.g. ref. [\horowitz ]). In the case of the heterotic string theory
one finds $g=2$ for all values of the charge $Q$ [\senbh ].
The same result holds in Einstein-Maxwell theory, but, generically,
$g$ is different than two; it is a function of the parameters characterizing
the black hole.

It is straightforward to generalize the result of
section 2 to the case of the heterotic string theory.
Eq. \dos\  can be applied to obtain expressions for the
left and right sectors of the bosonic string separately.
In addition we have the  Virasoro constraint which sets $n_L=n_R=n$
(more precisely, $n_L=n_R+ \tilde a $, the value of $\tilde a$
according to the sector, but for large the normal ordering constant
$\tilde a$ can be neglected). The discussion can be simplified
by setting $J<<n$ in eq. \dos , which means that $J$ is not on
a  Regge trajectory. It turns out that in the large $n$ case
we are considering this does not affect integrals over $J$. We have
$$
d_{n,J_L,J_R} \cong
{\rm const.}\ n^{-23/2} {e^{2(a_L+a_R)\sqrt{n}}\over
\cosh^2(a_LJ_L/2\sqrt{n})\cosh^2(a_RJ_R/2\sqrt{n}) }\ ,
\eqn\dhetero
$$
where $a_L=2\pi $ and $a_R=\sqrt{2}\pi $. Thus
$$
d_{n,J}= \int_ {-n}^n dJ_L d_{n,J_L,J-J_L} =\int_ {-n}^n
 dJ_R d_{n,J-J_R,J_R} \ .
\eqn\dtotal
$$

So let us consider a macroscopic four-dimensional rotating and
electrically charged black hole in the heterotic string theory and
let us choose the $z$-direction along the angular momentum, so that
$J_z=|J|, \ J_x=J_y=0 $.
Now let us study a thermal ensemble of strings with given mass,
charge and angular momentum. The charge $Q$ is the  quantum number
of a zero-mode operator and characterizes the Fock space vacuum.
As long as $Q<<m$,  it does not play any role in the following discussion.
If $Q=O(m)$ then the condition $n_L\cong n_R$ must be modified by the addition
of a term proportional to $Q^2$.
Let us compute the average magnetic dipole moment
$\langle \vec \mu \rangle \prop \langle \vec J_R \rangle $
in the thermal ensemble. Clearly $\langle J_R^x \rangle =
\langle J_R^y \rangle =0$, since for any configuration with $J^x_R$ there
is a configuration with $-J_R^x$.
The average gyromagnetic factor will be given by
$$
\langle g \rangle =2 {\langle J_R^z \rangle\over J}={\langle J_R \rangle\over
\langle J_R \rangle+\langle J_L \rangle}\ ,
\eqn\giro
$$
with
$$
\langle J_L \rangle ={1\over d_{n,J}}
\int _ {-n}^n dJ_L  J_L d_{n,J_L,J-J_L} \ ,\ \ \
\langle J_R \rangle ={1\over d_{n,J}}
\int _ {-n}^n dJ_R  J_R d_{n,J-J_R,J_R} \ .
\eqn\medias
$$
We have computed the integrals in eqs. \dtotal\ and \medias\
by numerical integration for different values of $J$ and $n$.
Three cases can be distinguished:

\n i) $\lim_{n\to\infty} {J\over\sqrt{n}}=0 $ . In this case we obtain
$\langle g\rangle \cong 1.27$ .

\n ii) $\lim_{n\to\infty} {J\over\sqrt{n}}=\infty$, for which we find
$\langle g\rangle =2$ .

\n iii) $\lim_{n\to\infty} {J\over\sqrt{n}}=\bar J= {\rm finite}$ . Then
$\langle g\rangle $ is a monotonically increasing function of
$\bar J$ which interpolates between 1.27 and 2.

\n As we will see, the correct scaling arises automatically from the
correspondence between level densities.

Because a string in the vicinity of the
horizon is described in terms of a Rindler geometry, in ref. [\susskind ] it
was argued that the black hole ADM mass $M$ must be related to the string
energy by $m\sim M^2$.
The Bekenstein-Hawking entropy for a rotating black hole
is given by $S=A/4G$,  with ($Q^2<<M$) [\senbh ]
$$
A=8\pi M^2G^2\bigg[ 1+\bigg(1-{J^2\over G^2M^4}\bigg)^{1/2}\bigg]\ .
\eqn\entropias
$$
On the other hand from eqs. \dhetero, \dtotal\  we see that, for large
$n$, $d_{n,J}$ can be written as
$$
d_{n,J}=e^{2(a_L+a_R)\sqrt{n} }f(J/\sqrt{n})\ .
\eqn\dnjei
$$
The function $f(J/\sqrt{n})$ can be explicitly obtained from eq. \dtotal\
for large $\bar J= J/\sqrt{n}$. It is of the form $f(\bar J)=c_0
e^{-a_R\bar J}+c_1 e^{-a_L\bar J}+...$.

{}From eqs. \entropias\ and \dnjei\ we see that correspondence between
heterotic string and black hole density level requires
$$
2(a_L+a_R)\sqrt n +\log f(J/\sqrt{n}) = M^2 G F(J/GM^2)
\eqn\corresp
$$
where
$$
F(J/GM^2)=2\pi \bigg[ 1+\bigg(1-{J^2\over G^2M^4}\bigg)^{1/2}\bigg]
$$
Eq. \corresp \ establishes a relationship between the string energy
and the black hole ADM mass as a function of the angular momentum.
Remarkably, this relation implies that
the only consistent scaling is $m\sim M^2 $.
Indeed, let us take the
limit $J\to\infty $ and $M\to\infty $ at $j\equiv J/M^2$ fixed.
In case i), we have
$\log f(0)={\rm finite}$, and we get from eq. \corresp\
$m={\rm const.}\  M^2+{\rm finite}$. This is in contradiction with
the assumption that $J/\sqrt n\to 0$ at $J/M^2$ fixed,
and therefore this case is not consistent with eq. \corresp .
Similarly, in case ii) we get $m\sim M^2+M^2/m$. Again, this
contradicts the assumption that $J/\sqrt n\to\infty$ at $J/M^2$ fixed.
Finally, the case iii) leads
to an equation of the form
$$
{m\over T_H}+\log f\big( 2J/m\sqrt\ap \big)=M^2 G F(J/M^2G) ,
\eqn\relacion
$$
where $T_H={1\over\pi\sqrt \ap}\big(1-{1\over \sqrt 2}\big)$. Eq. \relacion\
generalizes the relation $m=4\pi M^2 G T_H$ to the case $J\neq 0$.

Thus the correspondence between the level
density requires that $m\sim M^2$ and hence $J$ scales with $\sqrt n$.
As mentioned above, in this case the average gyromagnetic factor
is a function of $J/M^2G$ varying between $\sim 1.27$ and 2.
However, it should be noted that the calculation of $g$
is extremely sensitive to the scaling, since
logarithmic corrections to the scaling, such as $J\sim n^{1/2+\ee} ,
 \ \ee >0$ would lead to $g=2$, which is the value for rotating black holes
in heterotic string theory.

\bigskip

One of the authors (J.R.) wishes to thank S. Ferrara for a useful discussion
and for calling his attention to refs. [\carter , \ferrara ]. He also
thanks W. Fischler for very valuable remarks. L.S. would like to thank
V. Teledgi for an interesting conversation.

\refout
\vskip 2cm


\vfill\eject
\end